\documentclass[onecolumn,pra]{revtex4}
\usepackage{bm}
\usepackage{epsfig}
\usepackage{amsmath}
\begin{document}
\title{Multi-Channel Electron Transfer Reactions: An Analytically Solvable Model}
\author{Aniruddha Chakraborty \\
School of Basic Sciences, Indian Institute of Technology Mandi,\\
Mandi, Himachal Pradesh, 175001, India}
\date{\today }
\begin{abstract}
\noindent  We propose an analytical method for understanding the problem of multi-channel electron transfer reaction in solution, modeled by a particle undergoing diffusive motion under the influence of one donor and several acceptor potentials. The coupling between the donor potential and acceptor potentials are assumed to be represented by Dirac Delta functions. The diffusive motion in this paper is represented by the Smoluchowskii equation. Our solution requires the knowledge of the Laplace transform of the Green's function for the motion in all the uncoupled potentials. 
\end{abstract}
\maketitle
Understanding of electron transfer processes in condensed phase is very important in chemistry, physics, and biological sciences, for the experimentalists as well as theoreticians \cite{Marcus,MarcusRev,Zusman1,Zusman2,Zusman3,Wolynes1,Wolynes2,Hynes,Barbara1,Barbara2,Myers,Ivanov,Jortner,Cukier,Sumi,Maroncelli,Dutton1,Dutton2,Bagchi,AniJCP}. A large amount of research in this area has been dedicated in the understanding of the behavior of electron transfer reactions exhibited by donor-acceptor pairs in solutions. Multi-channel electron transfer in condensed phase is one of the very interesting problem to study. In general the quantum "jumps" of a high frequency vibrational mode can open several new point reaction sinks for electron transfer \cite{Bixon}, in contrast to the broadening effect by a low frequency mode \cite{BagchiBook}. The work of Jortner and Bixon \cite{Bixon} was based on a quantum mechanical treatment of the high frequency vibrational coordinate in place of classical one of Sumi and Marcus \cite{Sumi}. However, theoretical treatment of Jortner and Bixon \cite{Bixon} did not consider the dynamics of motion on the potential surfaces. In the following we propose a simple analytical method for understanding the problem of multi-channel electron transfer reaction in solution, modeled by a particle undergoing diffusive motion under the influence of one donor and several acceptor potentials explicitly.  A molecule (donor - acceptors) immersed in a polar solvent can be put on an electronically excited potential (represents the free energy of the donor surface) by the absorption of radiation. The molecule executes a walk on that potential, which may be considered as random as it is immersed in the polar solvent. As the molecule moves it may undergo non-radiative decay from certain regions of that potential to several potentials (represents the free energy of acceptor potentials). So the problem is to calculate the probability that the molecule will still be on the electronically excited donor potential after a finite time $t$.
We denote  the probability that the molecule would survive on the donor potential by $P_{d}(x,t)$. We also use $P^{(i)}_{a}(x,t)$ to denote the probability that the molecule would be found on the $i$-th acceptor potential. It is very usual to assume the motion on all the potentials to be one dimensional and diffusive, the relevent coordinate being denoted by $x$. It is also common to assume that the motion on all the potential energy surface is overdamped. Thus all the probability 
$P_{d}(x,t)$, $P^{(i)}_{a}(x,t)$s may be found at $x$ at the time $t$ obeys a modified Smoluchowskii equation.
\begin{eqnarray}
\frac{\partial P_d(x,t)}{\partial t} = {\cal L}_d P_d(x,t)+k_r P_d(x,t)- \sum_{i=1}^{N}k_i S_i(x) P^{(i)}_{a}(x,t) \\ \nonumber
\frac{\partial P^{(1)}_a(x,t)}{\partial t} = {\cal L}_1 P^{(1)}_a(x,t)+k_r P^{(1)}_a(x,t)- k_2 S_2(x) P_{d}(x,t) \\ \nonumber
\frac{\partial P^{(2)}_a(x,t)}{\partial t} = {\cal L}_2 P^{(2)}_a(x,t)+k_r P^{(2)}_a(x,t)- k_3 S_3(x) P_{d}(x,t) \\ \nonumber
............................................................................................. \\ \nonumber
..............................................................................................\\ \nonumber
\frac{\partial P^{(N)}_a(x,t)}{\partial t} ={\cal L}_N P^{(N)}_a(x,t)+k_r P^{(N)}_a(x,t) - k_{N} S_{N}(x) P_{d}(x,t). \nonumber
\end{eqnarray}
In the above 
\begin{equation}
{\cal L}_i= D \left[ \frac{\partial^2}{\partial x^2}+ \beta \frac{\partial}{\partial x} \frac{dV_i(x)}{dx}\right].
\end{equation}
$V_i(x)$ is the potential causing the drift of the particle, $S_{i}(x)$ is a position dependent sink function, $k_r$ is the rate of radiative decay and $k_0$ is the rate of electron transfer. We have taken $k_r$ to be independent of position. $D$ is the diffusion coefficient. Before we excite, the molecule is in the ground state, and as the solvent is at a finite temperature, its distribution over the coordinate $x$ is random. From this it undergoes Franck-Condon excitation to the excited state potential (donor). So, $x_0$ the initial position of the particle, on the excited state potential is random. We assume it to be given by the probability density $P^{0}_{d} (x_0)$. In the following we provide a general procedure for finding the exact analytical solution of Eq. (1). The Laplace transform ${\cal P}_i(x,s)=\int_{0}^{\infty} P_i(x,t)e^{-st} dt$ obeys
\begin{eqnarray}
[s-{\cal L}_d+k_r] {\cal P}_d(x,s)+\sum_{i=1}^{N}k_i S_i(x){\cal P}_i(x,s) = P^0_d(x_0) \\ \nonumber
[s-{\cal L}_1+k_r] {\cal P}^1_a(x,s)+ k_1 S_1(x) {\cal P}_d (x,s) = 0, \\ \nonumber
[s-{\cal L}_2+k_r] {\cal P}^2_a(x,s)+ k_2 S_2(x) {\cal P}_d (x,s) = 0, \\ \nonumber
........................................................... \\ \nonumber
............................................................\\ \nonumber
[s-{\cal L}_{N}+k_r] {\cal P}^{N}_a(x,s)+ k_{N} S_{N}(x) {\cal P}_d (x,s) = 0,
\end{eqnarray}
where $P^0_d(x_0)=P_d(x,0)$ is the initial distribution at the electronically excited  state (donor potential) and $P^{i}_a(x,0)=0$ is the initial distribution at all the acceptor potentials. In the following we will derive an analytical solution for this problem. We start with the simplest version of the problem {\it i.e.} donor potential and one acceptor potential. For this version of the problem Eq.(3) can be simplified as shown below.
\begin{eqnarray}
[s-{\cal L}_d+k_r] {\cal P}_d(x,s)+k_1 S_1(x){\cal P}^{(1)}_a(x,s) = P^0_d(x_0) \\ \nonumber
[s-{\cal L}_1+k_r] {\cal P}^{(1)}_a(x,s)+ k_1 S_1(x) {\cal P}_d (x,s) = 0.
\end{eqnarray}
So ${\cal P}^{(1)}_a(x,s)$ can be expressed as
\begin{equation}
{\cal P}^{(1)}_a(x,s)= [s-{\cal L}_1+k_r]^{-1} k_1 S_1(x) {\cal P}_d (x,s).
\end{equation}
Now we substitute ${\cal P}^{(1)}_a(x,s)$ to the first equation of Eq. (4) to get
\begin{equation}
[s-{\cal L}_d+k_r] {\cal P}_d(x,s)+{k_1}^2 S_1(x)[s-{\cal L}_1+k_r]^{-1} S_1(x) {\cal P}_d (x,s) = P^0_d(x_0)
\end{equation}
The above equation simplifies considerably if the coupling is assumed to be Dirac Delta function, which in operator notation may be written as  $S_1= \left| x_1 \left> \right < x_1 \right |$. The above equation may be written as
\begin{equation}
[s-{\cal L}_d+k_r] {\cal P}_d(x,s)+{k_1}^2 \delta(x - x_1)G^{0}_1(x_1,s;x_1){\cal P}_d (x,s) = P^0_d(x_0),
\end{equation}
where
\begin{equation}
G^{0}_1(x,s;x_0)=\left < x \left|[s-{\cal L}_1+ k_r ]^{-1}\right| x_0 \right>.
\end{equation}
Using the partition technique \cite{Lowdin}, solution of the Eq.(7) can be written as 
\begin{equation}
{\cal P}_{d \rightarrow 1}(x,s)=\int_{-\infty}^{\infty} dx_0 G^{0}_{d \rightarrow 1}(x,s;x_0)P^0_d(x_0),
\end{equation}
where $G^{0}_{d \rightarrow 1}(x,s;x_0)$ is the Green's function defined by the following equation
\begin{equation}
G^{0}_{d \rightarrow 1}(x,s;x_0)=\left < x \left|[s-{\cal L}_d+ k_r + {k_1}^2 G^{0}_1(x_1,s;x_1) S ]^{-1}\right| x_0 \right>,
\end{equation}
and corresponds to propagation of the particle starting from $x_0$ on the first acceptor potential in the absence of any coupling. Now we use the operator identity
\begin{equation}
[s-{\cal L}_d + k_r + {k_1}^2 G^{0}_1(x_1,s;x_1) S]^{-1}=[s-{\cal L}_d+ k_r]^{-1}-[s-{\cal L}_d+ k_r]^{-1}{k_1}^2 G^{0}_1(x_1,s;x_1) S [s-{\cal L}_d + k_r - {k_1}^2 G^{0}_1(x_1,s;x_1) S]^{-1}
\end{equation}
Inserting the resolution of identity $I=\int_{-\infty}^{\infty} dy \left|y \left> \right < y \right|$ in the second term of the above equation, we arrive at an equation which is very similar to Lippman-Schwinger equation.
\begin{equation}
G^{0}_{d \rightarrow 1}(x,s;x_0)=G^0_d(x,s;x_0) - {k_1}^2 G^0_d(x,s;x_1)G^0_1(x_1,s;x_1)G^{0}_{d \rightarrow 1}(x_1,s;x_0).
\end{equation}
where $G^0_d(x,s;x_0)=\left < x \left|[s-{\cal L}_d+k_r]^{-1}\right| x_0 \right>$ corresponds to the propagation of the particle on donor potential put initially at $x_0$, in the absence of any coupling, it is actually the Laplace Transform of $G^0_d(x,t;x_0)$, which is the probability that a particle starting at $x_0$ can be found at $x$ at time $t$. We now put $x=x_1$ in the above equation and solve for $G^0_{d \rightarrow a}(x_1,s;x_0)$ to get
\begin{equation}
G^0_{d \rightarrow 1}(x,s;x_0)=\frac{G^0_d(x,s;x_0)}{1 + {k_1}^2 G^0_d(x_1,s;x_1)G^0_1(x_1,s;x_1)}.
\end{equation}
This when substituted back into Eq. (12) gives
\begin{equation}
G^{0}_{d \rightarrow 1}(x,s;x_0)=G^0_d(x,s;x_0) - \frac{{k_1}^2 G^0_d(x,s;x_1)G^0_1(x_1,s;x_1)G^0_d(x_1,s;x_0)}{1+{k_1}^2 G^0_d(x_1,s;x_1)G^0_1(x_1,s;x_1)}.
\end{equation}
So if we know the analytical form of both $G^0_d(x,s;x_0)$ and $G^0_1(x,s;x_0)$, we can derive an analytical expression for $G^{0}_{d \rightarrow 1}(x,s;x_0)$. Now we consider the following case, the donor potential and two acceptor potentials, so that Eq. (4) will be modified to the following one
\begin{eqnarray}
[s-{\cal L}_d+k_r] {\cal P}_d(x,s)+k_1 S_1(x){\cal P}^{(1)}_a(x,s)+k_2 S_2(x){\cal P}^{(2)}_a(x,s) = P^0_d(x_0) \\ \nonumber
[s-{\cal L}_1+k_r] {\cal P}^{(1)}_a(x,s)+ k_1 S_1(x) {\cal P}_d (x,s) = 0 \\ \nonumber
[s-{\cal L}_2+k_r] {\cal P}^{(2)}_a(x,s)+ k_2 S_2(x) {\cal P}_d (x,s) = 0.
\end{eqnarray}
Now we express ${\cal P}^{(1)}_a(x,s)$ and ${\cal P}^{(2)}_a(x,s)$ in terms of ${\cal P}_d (x,s)$, in the first equation of Eq. (15) to get
\begin{equation}
[s-{\cal L}_d+k_r] {\cal P}_d(x,s)+{k_1}^2 S_1(x)[s-{\cal L}_1+k_r]^{-1} S_1(x) {\cal P}_d (x,s)+{k_2}^2 S_2(x)[s-{\cal L}_2+k_r]^{-1} S_2(x) {\cal P}_d (x,s) = P^0_d(x_0).
\end{equation}
The above equation simplifies considerably if the couplings are assumed to be Dirac Delta function, which in operator notation may be written as  $S_i= \left| x_i \left> \right < x_i \right |$. The above equation may be written as
\begin{equation}
[s-{\cal L}_d+k_r] {\cal P}_d(x,s)+{k_1}^2 \delta(x - x_1)G^{0}_1(x_1,s;x_1){\cal P}_d (x,s)+{k_2}^2 \delta(x - x_2)G^{0}_2(x_2,s;x_2){\cal P}_d (x,s) = P^0_d(x_0),
\end{equation}
where
\begin{equation}
G^{0}_2(x,s;x_0)=\left < x \left|[s-{\cal L}_2+ k_r ]^{-1}\right| x_0 \right>.
\end{equation}
Using the partition technique \cite{Lowdin}, solution of this equation can be written as 
\begin{equation}
{\cal P}_{d \rightarrow 2}(x,s)=\int_{-\infty}^{\infty} dx_0 G^{0}_{d \rightarrow 2}(x,s;x_0)P^0_d(x_0),
\end{equation}
where $G^{0}_{d \rightarrow 2}(x,s;x_0)$ can be derived from $G^{0}_{d \rightarrow 1}(x,s;x_0)$ and $G^0_2(x_1,s;x_1)$ using the same method as we have used in deriving Eq.(14), with the assumtion that the second potential in coupled to first one via Dirac delta function at $x_2$.
\begin{equation}
G^{0}_{d \rightarrow 2}(x,s;x_0)=G^0_{d \rightarrow 1}(x,s;x_0) - \frac{{k_2}^2 G^0_{d \rightarrow 1}(x,s;x_2)G^0_2(x_2,s;x_2)G^0_{d \rightarrow 1}(x_2,s;x_0)}{1+{k_2}^2 G^0_{d \rightarrow 1}(x_2,s;x_2)G^0_2(x_2,s;x_2)},
\end{equation}
where $G^0_2(x,s;x_0)=\left < x \left|[s-{\cal L}_2+k_r]^{-1}\right| x_0 \right>$ corresponds to the propagation of the particle on the second potential put initially at $x_0$, in the absence of any coupling. So one can use the similar procedure to deal with``N" channel problem, the N-channel generalisation of Eq. (17) is given below
\begin{equation}
[s-{\cal L}_d+k_r] {\cal P}_d(x,s)+ \sum_{\epsilon=1}^{N}{k_\epsilon}^2 \delta(x - x_\epsilon)G^{0}_\epsilon(x_\epsilon,s;x_\epsilon){\cal P}_d (x,s)= P^0_d(x_0),
\end{equation}
where $G^0_\epsilon(x,s;x_0)=\left < x \left|[s-{\cal L}_\epsilon+k_r]^{-1}\right| x_0 \right>$ corresponds to the propagation of the particle on the $\epsilon$-th potential put initially at $x_0$, in the absence of any coupling.
\begin{equation}
G^{0}_{d \rightarrow N}(x,s;x_0)=G^0_{d \rightarrow N-1}(x,s;x_0) - \frac{{k_N}^2 G^0_{d \rightarrow N-1}(x,s;x_N)G^0_N(x_N,s;x_N)G^0_{d \rightarrow N-1}(x_N,s;x_0)}{1+{k_N}^2 G^0_{d \rightarrow N-1}(x_N,s;x_N)G^0_N(x_N,s;x_N)},
\end{equation}
where $G^0_a(x,s;x_0)=\left < x \left|[s-{\cal L}_a+k_r]^{-1}\right| x_0 \right>$ corresponds to the propagation of the particle on the acceptor potential put initially at $x_0$, in the absence of any coupling. Using this Green's function one can calculate the corresponding ${\cal P}_{d \rightarrow N}(x,s)$ explicitely using the following equation. 
\begin{equation}
{\cal P}_{d \rightarrow N}(x,s)=\int_{-\infty}^{\infty} dx_0 G^{0}_{d \rightarrow N}(x,s;x_0)P^0_d(x_0),
\end{equation}
In the following, we consider an analytically solvable continuum model. Here we start with the $N$-channel generalization of Eq.(16).
\begin{equation}
[s-{\cal L}_d+k_r] {\cal P}_d(x,s)+ \sum_{\epsilon=1}^{N}{k_{\epsilon}(x)}^2 S_{\epsilon}(x)[s-{\cal L}_{\epsilon}+k_r]^{-1} S_{\epsilon}(x) {\cal P}_d (x,s) = P^0_d(x_0).
\end{equation}
Now we take $S_{\epsilon}(x)= \delta(x-x_{\epsilon})$, so the above expression becomes
\begin{equation}
[s-{\cal L}_d+k_r] {\cal P}_d(x,s)+ \sum_{\epsilon=1}^{N}{k_{\epsilon}(x)}^2 \delta(x-x_{\epsilon}) G^{0}_\epsilon(x_\epsilon,s;x_\epsilon) {\cal P}_d (x,s) = P^0_d(x_0).
\end{equation}
In the above equation, $\epsilon$ takes discrete values. The solution of this equation may be expressed in terms of the Greens function $G(x,s;x_0)$.
\begin{equation}
\left([s-{\cal L}_d+k_r]+ \sum_{\epsilon=1}^{N}{k_{\epsilon}(x)}^2 \delta(x-x_{\epsilon}) G^{0}_\epsilon(x_\epsilon,s;x_\epsilon)\right) G^{0}_{d \rightarrow N}(x,s;x_0) = \delta(x-x_0).
\end{equation}
Here
\begin{equation}
G^{0}_{d \rightarrow N}(x,s;x_0)=\left < x \left|\left([s-{\cal L}_d+k_r]+ \sum_{\epsilon=1}^{N}{k_{\epsilon}(x)}^2 \delta(x-x_{\epsilon}) G^{0}_\epsilon(x_\epsilon,s;x_\epsilon)\right)^{-1}\right| x_0 \right>.
\end{equation}
Also
\begin{equation}
{\cal P}_{d \rightarrow N}(x,s)=\int_{-\infty}^{\infty} dx_0 G^{0}_{d \rightarrow N}(x,s;x_0)P^0_d(x_0).
\end{equation}
In the continuum limit (here $\epsilon$ changes continuously), the expression for $G^{0}_{d \rightarrow N}(x,s;x_0)$ can be written
\begin{equation}
G^{0}_{d \rightarrow N}(x,s;x_0)=\left < x \left|\left([s-{\cal L}_d+k_r]+ \int_0^\infty d{\epsilon}{k(x, \epsilon)}^2 \delta(x-x[\epsilon]) G^{0}(x[\epsilon],s;x[\epsilon],\epsilon)\right)^{-1}\right| x_0 \right>.
\end{equation}
Now we consider a special case, where all states couple at one point to the first one, {\it i.e.,} $s(\epsilon)=a$. The expression for Greens function now become
\begin{equation}
G^{0}_{d \rightarrow N}(x,s;x_0)=\left < x \left|\left([s-{\cal L}_d+k_r]+ \int_0^\infty d{\epsilon}{k(x, \epsilon)}^2 \delta(x-a) G^{0}(a,s;a,\epsilon)\right)^{-1}\right| x_0 \right>.
\end{equation}
The expression we have derived above is very general and applicable to any potential. We can write the above equation in a very simplified form
\begin{equation}
G^{0}_{d \rightarrow N}(x,s;x_0)=\left < x \left|\left([s-{\cal L}_d+k_r]+  \delta(x-a) V(x,s)\right)^{-1}\right| x_0 \right>,
\end{equation}
where
\begin{equation}
V(x,s)=\int_0^\infty d{\epsilon}{k(x, \epsilon)}^2 G^{0}(a,s;a,\epsilon).
\end{equation}
In the following, we discuss the case where one arbitrary potential is coupled to a continuum (in energy) of constant potentials at a point. So the Greens function for an uncoupled potential is givem by
\begin{equation}
G^{0}(a,s;a,\epsilon)=\frac{1}{2\sqrt{D (i s - \epsilon)}}.
\end{equation}
In the above expression $\epsilon$ varies continuously from $0$ to $\infty$. It is interesting to note that the analytical form of $V(x,s)$ depends on the functional form of $k(x, \epsilon)$. For further calculation we take
\begin{equation}
{k(x,\epsilon)}^2= 2 a \sqrt{D (i s - \epsilon)} \exp{(- \frac{\epsilon}{s})}.
\end{equation} 
So,
\begin{equation}
V(x,s)= a \int_0^\infty d \epsilon \exp{(- \frac{\epsilon}{s})} = a s.
\end{equation}
Now Eq. (31) becomes 
\begin{equation}
G^{0}_{d \rightarrow N}(x,s;x_0)=\left < x \left|\left([s-{\cal L}_d+k_r]+  a s \delta(x-a) \right)^{-1}\right| x_0 \right>.
\end{equation}
The above expression can be evaluated exactly. Once we have an expression for $G^{0}_{d \rightarrow N}(x,s;x_0)$, we can use Eq.(23) to derive an expression for ${\cal P}_{d \rightarrow N}(x,s)$ easily and the expression for ${\cal P}_{d \rightarrow N}(x,s)$ can be used to calculate survival probability at the donor potential and the corresponding rate constants. The survival probability at the donor potential may be defined as follows 
\begin{equation}
P_{d \rightarrow N}(t) = \int_{-\infty}^{\infty} dx P_{d\rightarrow N}(x,t).
\end{equation}
It is possible to evaluate Laplace Transform  ${\cal P}_{d\rightarrow N}(s)$ of $P_{d\rightarrow N}(t)$ directly. 
${\cal P}_{d\rightarrow N} (s)$ is defined in terms of ${\cal P_{d\rightarrow N}}(x,s)$ by the following equation,
\begin{equation}
{\cal P}_{d \rightarrow N}(s) = \int_{-\infty}^{\infty} dx {\cal P}_{d \rightarrow N}(x,s).
\end{equation}
So now one can use Eq.(23) to calculate ${\cal P}_{d \rightarrow N} (s)$. The average and long time rate constants can be found from ${\cal P}_{d \rightarrow a}(s)$ \cite{Kls2} as given below
\begin{equation}
k^{-1}_{1}={\cal P}_{d \rightarrow N}(0).
\end{equation}
Also $k_{L}=$ negative of the pole of ${\cal P}_{d \rightarrow N}(s)$, closest to the origin. \par
The method we have discussed here can also be applied to the case where $S$ is a nonlocal operator, and may be represented by  $S=\left|f\right>k_0\left <g \right|$, where $f$ and $g$ are arbitrary acceptable functions. Choosing both of them to be Gaussian will be an interesting improvement over the current model. $S$ can also be a linear combination of such operators.

\end{document}